\documentclass[twocolumn,prc,superscriptaddress,amsmath,amssymb]{revtex4-1}%
\usepackage{epsfig,dsfont,amssymb,amsmath,amsthm,amsfonts,amsbsy,mathrsfs}
\usepackage{graphicx}
\usepackage{amsmath}
\usepackage{amssymb}%
\usepackage{booktabs}
\usepackage{multirow}
\usepackage{lineno}
\usepackage{dcolumn}
\usepackage{bm}
\usepackage{color}
\usepackage{float}

\usepackage[colorlinks,
           bookmarks=true,
           linkcolor=blue,
           urlcolor=blue,
            anchorcolor=black,
            citecolor=blue
            ]{hyperref}

\usepackage[all]{hypcap}

\begin{document}
	\title{Intermittency of charged particles in the hybrid UrQMD+CMC model at energies available at the BNL Relativistic Heavy Ion Collider}

\author{Jin Wu}
\affiliation{Key Laboratory of Quark and Lepton Physics (MOE) and Institute of Particle Physics, \\
Central China Normal University, Wuhan 430079, China}

\author{Zhiming Li}
\email{lizm@mail.ccnu.edu.cn}
\affiliation{Key Laboratory of Quark and Lepton Physics (MOE) and Institute of Particle Physics, \\
Central China Normal University, Wuhan 430079, China}

\author{Xiaofeng Luo}
\email{xfluo@mail.ccnu.edu.cn}
\affiliation{Key Laboratory of Quark and Lepton Physics (MOE) and Institute of Particle Physics, \\
Central China Normal University, Wuhan 430079, China}

\author{Mingmei Xu}
\affiliation{Key Laboratory of Quark and Lepton Physics (MOE) and Institute of Particle Physics, \\
Central China Normal University, Wuhan 430079, China}

\author{Yuanfang Wu}
\email{wuyf@mail.ccnu.edu.cn}
\affiliation{Key Laboratory of Quark and Lepton Physics (MOE) and Institute of Particle Physics, \\
Central China Normal University, Wuhan 430079, China}

\begin{abstract}
Within the framework of intermittency analysis, a search for critical fluctuations is ongoing to locate the possible critical point in the quantum chromodynamics phase diagram. In this study, self-similar critical fluctuations from a critical Monte Carlo (CMC) model have been incorporated into the cascade ultrarelativistic quantum molecular dynamics (UrQMD) model. This hybrid UrQMD+CMC model exhibits a clear power-law behavior of scaled factorial moment for charged particles in Au+Au collisions at $\sqrt{s_\mathrm{NN}}$ = 7.7-200 GeV. By comparing the UrQMD+CMC model results with those from the STAR experiment, it is found that the value of a calculated scaling exponent falls in the range of the experimental measurement when 1-2 \% signal of intermittency fluctuations is added into the UrQMD sample. 
\end{abstract}
	
\maketitle
	
\section{Introduction}
One of the major goals in heavy-ion collisions is to locate the critical point in the phase diagram of strongly interacting matter predicted by quantum chromodynamics (QCD)~\cite{QCDReport,STARPRLMoment,MomentNucl,CEP1,CEP2,luo2022properties}. Experiments at BNL Relativistic Heavy Ion Collider (RHIC)~\cite{QCDReport,STARPRLMoment,STARPRLMoment3GeV,besII} and CERN Super
Proton Synchrotron (SPS)~\cite{NA61Coll,NA61QM2019} are ongoing to search for possible signals by measuring nuclear collisions at various energies. A generic feature of the critical point is the divergence of correlation length in its vicinity, leading to the system becomes scale-invariant and self-similar~\cite{invariant,BialasPLB,invariant2}. In analog to the critical opalescence observed in the critical system in quantum electrodynamics (QED), the fractal and self-similar geometry of the matter near the QCD critical point will give rise to giant local density fluctuation that manifests itself as critical intermittency in heavy-ion collisions~\cite{AntoniouPRL, AntoniouPRD}. Intermittency is able to be revealed in transverse momentum space as a power-law scaling of a scaled factorial moment (SFM)~\cite{AntoniouPRL, AntoniouPRD,AntoniouPRC,BialasPLB}. The strength of intermittency can be quantified by the intermittency index extracted from the power-law scaling of SFMs on the number of partitioned cells in momentum space~\cite{AntoniouPRL, AntoniouPRD} or by the scaling exponents obtained from the power-law of higher-order SFMs on the second-order one~\cite{GLPRL,GLPRD,GLPRC}.

During the last decades, experimental explorations on signatures of critical fluctuations expected for the QCD critical point have been performed by using the intermittency analysis at various system sizes and energies in heavy-ion collisions~\cite{NA49EPJC,NA49PRC,NA61universe,NA61Davis,STARIntermittency,QMSTARIntermittency}. The second-order SFM of protons has been found to obey a strong power-law behavior in Si+Si collisions at 158$A$ GeV from the NA49 experiment~\cite{NA49EPJC}. Recent preliminary result from the NA61 Collaboration illustrates that there seems no indication of a power-law intermittency in central Ar+Sc collisions at 150$A$ GeV/$c$ or in central Pb+Pb collisions at 30$A$ GeV/$c$~\cite{CPODNA61Intermittency,NA61CFM,QMNA61Intermittency}. The STAR experiment at RHIC has presented the preliminary results on intermittency of charged particles in Au+Au collisions at $\sqrt{s_\mathrm{NN}}$ =  7.7-200 GeV~\cite{STARIntermittency,QMSTARIntermittency}. It is shown that a strict power-law scaling cannot be satisfied although the values of SFMs are rising with increasing number of partitioned cells in momentum space. However, the higher-order SFMs with respect to the second-order one follow good power-law behaviors. Furthermore, a measured scaling exponent exhibits a non-monotonic behavior on collision energy with a possible minimum around $\sqrt{s_\mathrm{NN}}$ =  20-30 GeV which needs to be understood with more theoretical and model inputs.  

In the meantime, various model studies have been conducted to try to understand the measured intermittency in experiments~\cite{CMCPLB,UrQMDLi,RefIntermittencyPRC,PYTHIAPRC,SGEPJA,SBEPJP,AMPTnu,IJMPE}. An overview of the results can be found in Ref.~\cite{LiOverview}. However, none of the models in the market can describe the latest intermittency measurement in the STAR experiment and therefore warrants further investigations. Among these models, the ultrarelativistic quantum molecular dynamics (UrQMD) is the one that can well simulate the dynamics of evolution in $A+A$ collisions and successfully describes several experimental results~\cite{MBUrQMD,HPUrQMD,ELUrQMD,HussainUrQMD}. This cascade model has been proven to be appropriate for a background study in the intermittency analysis since no critical self-similar mechanism is implemented in it~\cite{RefIntermittencyPRC,UrQMDLi}. On the other hand, a critical Monte Carlo (CMC) model can easily simulate critical intermittency driven by self-similar density fluctuations~\cite{AntoniouPRL,CMCPLB,NA49EPJC}. But it can only produce scale-invariant multiplicity distributions in momentum space and does not include evolution of the system or background effects in heavy-ion collisions. Therefore, it is meaningful to combine these two models together to get a hybrid UrQMD+CMC one. In the hybrid model, the self-similar density fluctuations generated by the CMC simulation are incorporated into the final-state multiplicity distributions in the UrQMD event sample. We will use this hybrid model to study intermittency at RHIC beam energy scan (BES) energies and try to understand the STAR experimentally measured results.

The paper is organized as follows. We introduce the method of intermittency analysis by using SFMs in Sec. II.  In Sec. III, we describe the cascade UrQMD model and the directly calculated SFMs of charged particles, including $p$, $\bar{p}$, $K^{\pm}$, and $\pi^{\pm}$, in Au+Au collisions at $\sqrt{s_\mathrm{NN}}$ = 7.7-200 GeV. An introduction of the CMC model and the study of intermittency in this model are presented in Sec. IV. The results of the measured SFMs and scaling exponents in the hybrid UrQMD+CMC model are discussed in Sec. V. A summary and outlook of the work is given in Sec. VI.

\section{Intermittency Analysis Method}
In high energy experiments, the intermittency driven by self-similar density fluctuations can be measured by calculating the SFMs of final state particles in momentum space~\cite{AntoniouPRL,NA49EPJC}. For this purpose, the available region of momentum space is partitioned into equal-size cells, and the $q$th-order SFM, $F_{q}(M)$, is defined as~\cite{AntoniouPRL,NA49EPJC}:
\begin{equation}
F_{q}(M)=\frac{\langle\frac{1}{M^{D}}\sum_{i=1}^{M^{D}}n_{i}(n_{i}-1)\cdots(n_{i}-q+1)\rangle}{\langle\frac{1}{M^{D}}\sum_{i=1}^{M^{D}}n_{i}\rangle^{q}},
 \label{Eq:FM}
\end{equation}

\noindent where $M^{D}$ is the number of cells in $D$-dimensional momentum space, $n_{i}$ is the measured multiplicity in the $i$th cell, and the angular bracket denotes an average over all the events.

If the system is near the QCD critical point, the SFMs are expected to obey a power-law scaling behavior with $M^{D}$ when $M$ is large enough:~\cite{AntoniouPRL, AntoniouPRC, NA49EPJC},
\begin{equation}
F_{q}(M)\sim (M^{D})^{\phi_{q}}, M\gg 1.
 \label{Eq:PowerLaw}
\end{equation}

\noindent Here $\phi_{q}$ is the intermittency index that characterizes the strength of intermittency. The power-law scaling of Eq.~\eqref{Eq:PowerLaw} is referred as $F_{q}(M)/M$ scaling. If critical fluctuations can survive the evolution of heavy-ion collision system, the critical $\phi_{2}$ is predicted to be $\frac{5}{6}$~\cite{AntoniouPRL} for baryon density and $\frac{2}{3}$ for $\sigma$ condensate~\cite{NGNPA2005} based on calculations from a critical equation of state belonging to the 3D Ising universality class.

The other promising power-law behavior that describes relationship between higher-order $F_{q}(M)$ and the second-order $F_{2}(M)$ is defined as~\cite{GLPRL, GLPRD, OchsPLB1988, Ochs1990ZPC}:
\begin{equation}
F_{q}(M)\propto F_{2}(M)^{\beta_{q}}, M\gg 1. 
 \label{Eq:FqF2scaing}
\end{equation}

\noindent We refer to the power-law scaling of Eq.~\eqref{Eq:FqF2scaing} as $F_{q}(M)/F_{2}(M)$ scaling. According to the Ginzburg-Landau (GL) description of a critical system, the $F_{q}(M)/M$ scaling in Eq.~\eqref{Eq:PowerLaw} may be washed out and thus is hard to be observed in real experiment since its value depends on particular critical parameters which would vary with temperature in the dynamical evolution of the collision system~\cite{GLPRL,GLPRD}. However, $F_{q}(M)/F_{2}(M)$ scaling in Eq.~\eqref{Eq:FqF2scaing} is independent on those parameters.

A scaling exponent ($\nu$) which describes the general consequences of phase transition, independent on precise values of critical parameters, is given by~\cite{GLPRL,GLPRD,GLPRC,nuAPLB}:
\begin{equation}
\beta_{q} \propto (q-1)^{\nu}. 
 \label{Eq:betaqscaing}
\end{equation}

It is suggested that the energy dependence of $\nu$ could shed light on the search of the QCD critical point~\cite{STARIntermittency}. The critical value of $\nu$ is 1.304 from the GL theory~\cite{GLPRL} and 1.0 from the Ising model~\cite{GLPRD,2DIsing} for the second-order phase transition. In Monte-Carlo calculations, the value of $\nu$ equals to 1.94 from (a multiphase transport) AMPT model~\cite{AMPTnu} and 1.824 from the HIJING model~\cite{nuAPLB}. 

To subtract background contributions in the calculation of SFMs, a correlator, $\Delta F_{q}(M)$, is defined in terms of SFMs calculated in the original events and in the corresponding mixed events ~\cite{NA49EPJC,NA49PRC,NA61universe,STARIntermittency}:
\begin{equation}
 \Delta F_{q}(M)=F_{q}(M)^{data}-F_{q}(M)^{mix}.
 \label{Eq:DeltaFq}
\end{equation}

 \begin{figure*}
     \centering
     \includegraphics[scale=0.91]{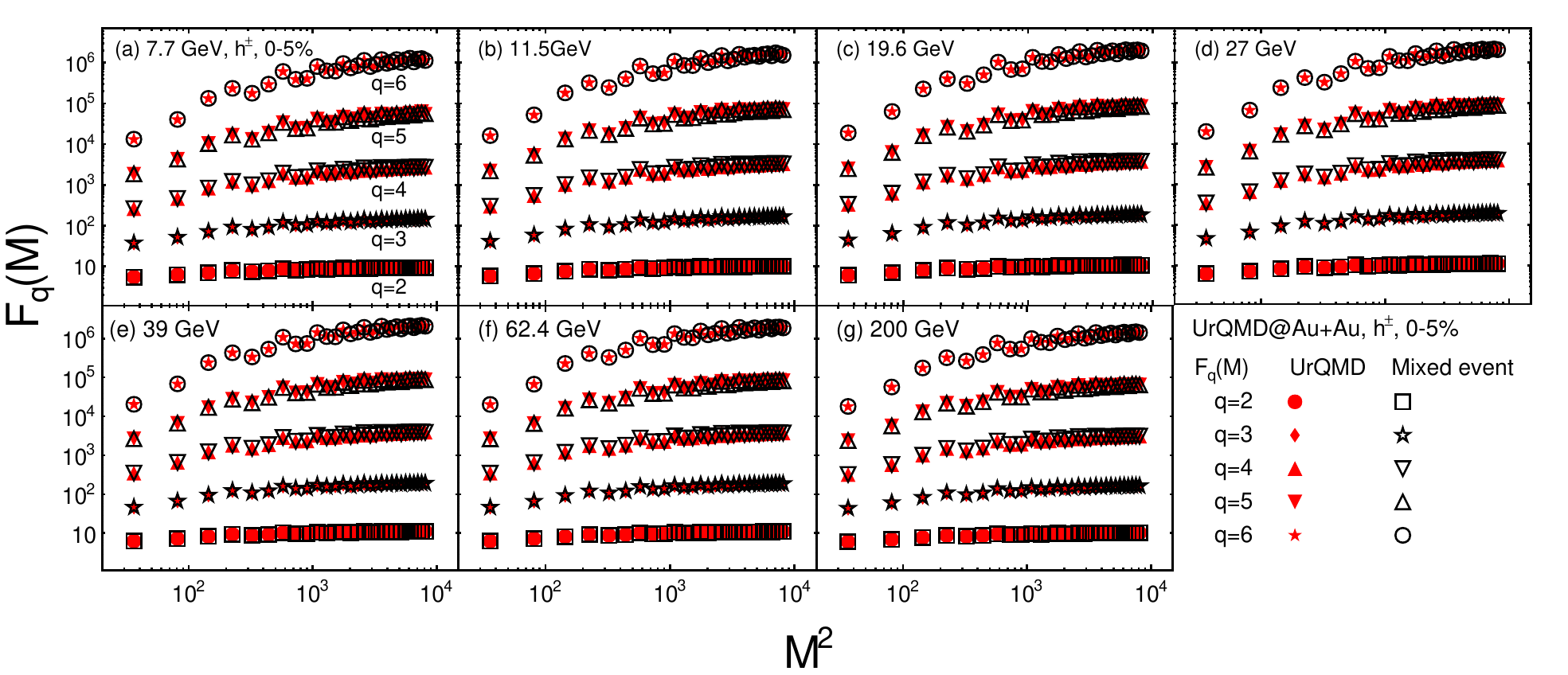}
      \caption{The scaled factorial moments, $F_{q}(M)$ (up to sixth order), as a function of number of cells ($M^{2}$) of charged particles in the most central (0-5\%) Au+Au collisions at $\sqrt{s_\mathrm{NN}}$ = 7.7-200 GeV from the UrQMD model in a double-logarithmic scale. Red (black) marks represent $F_{q}(M)$ of UrQMD data (mixed events), respectively. Statistical uncertainties are obtained from the Bootstrap method and are smaller than the maker size.}
     \label{Fig:FqmixFqUrQMD}
     \end{figure*}
     
\noindent The mixed event sample is constructed by randomly selecting particles from the original generated one, and each particle in a mixed event is chosen from different original events.  

\section{Intermittency of SFMs in the Cascade UrQMD Model}       
In high energy collisions, the UrQMD model has been widely and successfully applied to simulate $p+p$, $p+A$, and $A+A$ interactions~\cite{MBUrQMD,SABassUrQMD,HPUrQMD}. It is a microscopic transport approach which treats the covariant propagation of all hadrons as classical trajectories combined with stochastic binary scatterings, the excitation and fragmentation of color strings, and decay of hadronic resonances~\cite{MBUrQMD}. The model incorporates baryon-baryon, meson-baryon and meson-meson interactions with collision terms including more than 50 baryon and 45 meson species, and all particles can be produced in hadron-hadron collisions. Conservation law of electric charge and baryon number are taken into account in the model~\cite{MBPRCUrQMD,BurauUrQMD}. It can reproduce the cross-section of hadronic reactions, and successfully describe yields and momentum spectra of various particles in $A+A$ collisions~\cite{MBUrQMD,ELUrQMD}. The UrQMD is a well-designed transport model for simulations with the entire available range of energies from Schwerionen Synchrotron at GSI Darmstadt (SIS) energy ($\sqrt{s_\mathrm{NN}} \approx$ 2 GeV) to the top RHIC energy ($\sqrt{s_\mathrm{NN}}$ = 200 GeV). More details about the model can be found in Refs~\cite{MBUrQMD,SABassUrQMD,UrQMDPetersen}. 

The UrQMD model is a suitable simulator to estimate non-critical contributions from the hadronic phase as well as the associated physics processes since there is no phase transition to QGP state in the simulation. In this work, we use the cascade UrQMD model (version 3.4) to generate event samples in Au+Au collisions at RHIC energies. The corresponding event statistics are 1.54, 1.17, 1.15, 1.25, 1.20, 1.30, 0.5$\times 10^{6}$ at $\sqrt{s_\mathrm{NN}}$ = 7.7, 11.5, 19.6, 27, 39, 62.4, 200 GeV, respectively.

In our analysis, we apply the same analysis techniques and kinematic cuts as those used in the STAR experiment~\cite{STARIntermittency,QMSTARIntermittency}. Charged particles including proton ($p$), anti-proton ($\bar{p}$), kaons ($K^{\pm}$) and pions ($\pi^{\pm}$) are selected within pseudo-rapidity window ($\mid \eta \mid <0.5$), $p_{T}$ window ($0.2<p_{T}<1.6$ GeV/$c$) for $K^{\pm}$ and $\pi^{\pm}$, and ($0.4<p_{T}<2.0$ GeV/$c$) for $p$ and $\bar{p}$. To avoid auto-correlation effects, the centrality is determined from uncorrected charged particles within $0.5<\mid \eta \mid <1$, which is chosen to be beyond the analysis window $\mid\eta\mid<0.5$. Two dimensional transverse momentum space of $p_{x}$ and $p_{y}$ are partitioned into $M^{2}$ equal-size cells to calculate $F_{q}(M)$ with $M^{2}$ varying from 1 to $100^{2}$. The statistical error is estimated by the bootstrap method~\cite{RefBootstrap}.

In Fig.~\ref{Fig:FqmixFqUrQMD}, we show $F_{q}(M)$ of UrQMD data (red marks) and the corresponding mixed events (black marks) of charged particles, as a function of $M^{2}$ in the most central (0-5\%) Au+Au collisions at $\sqrt{s_\mathrm{NN}}$ = 7.7-200 GeV. $F_{q}(M)$ of UrQMD and associated mixed events are calculated up to the sixth order. It is found that $F_{q}(M)^{UrQMD}$ are almost overlapped with $F_{q}(M)^{mix}$, which leads to the correlator $\Delta F_{q}(M)\approx 0$. It implies that the magnitude of $F_{q}(M)^{UrQMD}$ are dominated by non-critical background contributions from the cascade UrQMD model. There should be no intermittency in this model since it does not incorporate any self-similar local density fluctuations. In contrast to the UrQMD model, $F_{q}(M)^{data}$ ($q=2-6$) are larger than $F_{q}(M)^{mix}$ and thus $\Delta F_{q}(M)$ increase with increasing $M^{2}$ at RHIC energies from the preliminary results of the STAR experiment~\cite{STARIntermittency,QMSTARIntermittency}.

     \begin{figure*}
     \centering 
     \includegraphics[scale=0.91]{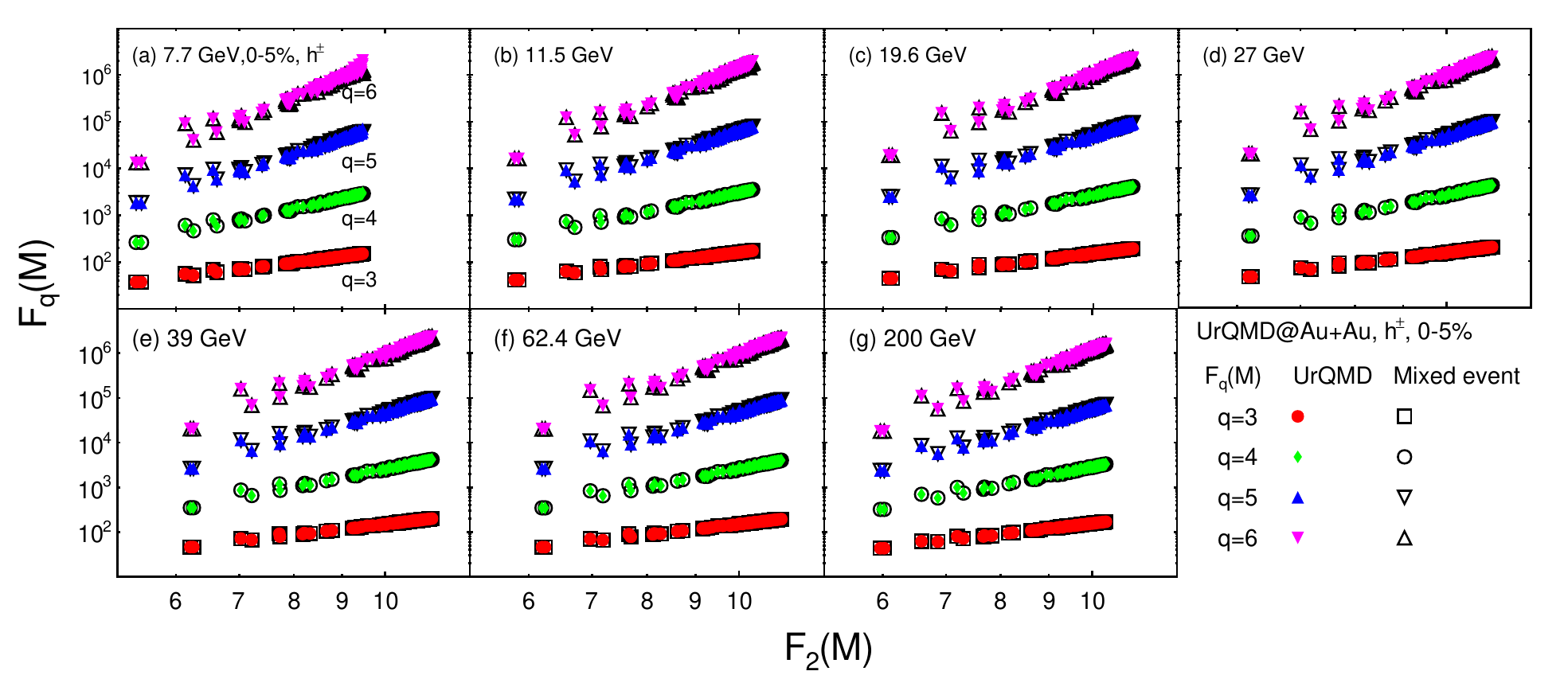}
      \caption{The higher-order $F_{q}(M)$ (q=3-6) of charged particles as a function of $F_{2}(M)$ in the most central (0-5\%) Au+Au collisions at $\sqrt{s_\mathrm{NN}}$ = 7.7-200 GeV from the UrQMD model in a double-logarithmic scale. Solid (open) marks represent $F_{q}(M)$ of UrQMD data (mixed events), respectively.}
     \label{Fig:FqF2ScalingUrQMD}
     \end{figure*}
     
We then investigate the $F_{q}(M)/F_{2}(M)$ scaling as introduced in Eq.~\eqref{Eq:FqF2scaing}. The solid symbols in Fig.~\ref{Fig:FqF2ScalingUrQMD} illustrate the higher-order $F_{q}(M)^{UrQMD}$ ($q$=3-6) of charged particles as a function of $F_{2}(M)^{UrQMD}$ in the most central (0-5\%) Au+Au collisions at $\sqrt{s_\mathrm{NN}}$ = 7.7-200 GeV. It seems that $F_{q}(M)^{UrQMD}$ ($q$=3-6) exhibit clear power-law scaling with $F_{2}(M)^{UrQMD}$. Whereas, the corresponding open symbols for the mixed events agree well with the UrQMD results. It means that background effects dominate the observed $F_{q}(M)/F_{2}(M)$ scaling in the cascade UrQMD samples. This scaling will be vanished if the background effects are subtracted from the UrQMD results by the mixed event method.

     \begin{figure*}
      \centering 
     \includegraphics[scale=0.91]{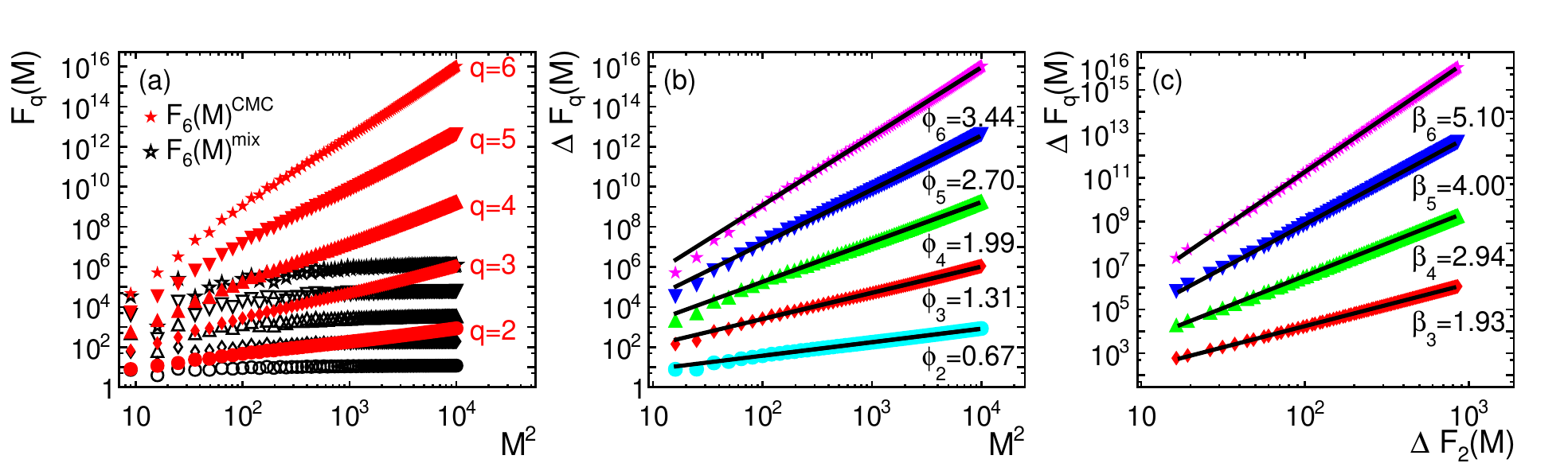}
      \caption{(a) The scaled factorial moments as a function of number of divided cells in the CMC (solid red symbols) and mixed event (open black ones) samples in a  double-logarithmic scale. (b) The correlator $\Delta F_{q}(M)$ (q=2-6) as a function of $M^{2}$ from the CMC model. The solid black lines are the fitting according to the power-law relation of Eq.~\eqref{Eq:PowerLaw}. (c) The higher-order $\Delta F_{q}(M)$ (q=3-6) as a function of $\Delta F_{2}(M)$. The solid black lines are the fitting according to Eq.~\eqref{Eq:FqF2scaing}.}
     \label{Fig:CMCIntermittency}
     \end{figure*}
  
\section{Intermittency of SFMs in the CMC Model}
To simulate critical events incorporating self-similar local density fluctuations, the CMC model~\cite{AntoniouPRL,CMCPLB,AntoniouNPA2020} is used to generate a series of particles of which any two-particle correlation follows a self-similar geometry. The momentum profiles of final state particles are produced by the algorithm of Levy random walk which requires the probability density $\rho(p)$ between two adjacent walks follows:
\begin{equation}
\rho(p)=\frac{\mu p_{\rm min}^{\mu}}{1-(p_{\rm min}/p_{\rm max})^{\mu}}p^{-1-\mu},
 \label{Eq:probLevy}
\end{equation}

\noindent where $\mu$ is the Levy exponent directly connected to the intermittency index, $p$ denotes the relative momentum of two adjacent particles in one-dimensional space, $p_{\rm min}$ and $p_{\rm max}$ are the minimum and maximum values of $p$, respectively. The parameters in Eq.~\eqref{Eq:probLevy} are set to be $\mu = 1/6$ and $p_{\rm min}/p_{\rm max} = 10^{-7}$ for critical system belonging to the 3D-Ising universality class~\cite{AntoniouPRL}. More details about the implementation of the CMC model can be found in Refs~\cite{AntoniouPRL,CMCPLB, PRELevy}.

In Fig.~\ref{Fig:CMCIntermittency} (a), the red symbols show the $F_{q}(M)/M$ scaling from the CMC sample which incorporates the same statistics, mean multiplicity and $p_{T}$ distributions as those in the UrQMD sample at $\sqrt{s_\mathrm{NN}}$= 19.6 GeV. $F_{q}(M)^{CMC}$ of all orders are found to rise with increasing $M^{2}$. The corresponding open black symbols are the results from the mixed events. we observe that $F_{q}(M)^{CMC}$ ($q$=2-6) are clearly larger than $F_{q}(M)^{mix}$, especially in large $M^{2}$ regions. After subtracting background by using the mixed event method, Fig.~\ref{Fig:CMCIntermittency} (b) shows the correlator $\Delta F_{q}(M)$  as a function of $M^{2}$. A good scaling behavior is satisfied for each order of $\Delta F_{q}(M)$, i.e. $\Delta F_{q}(M)/M$ scaling is observed in the CMC model. Fig.~\ref{Fig:CMCIntermittency} (c) presents $\Delta F_{q}(M)$ ($q$=3-6) as a function of $\Delta F_{2}(M)$. It is found that the correlators $\Delta F_{q}(M)$ follow strict $\Delta F_{q}(M)/\Delta F_{2}(M)$ scaling as illustrated in Eq.~\eqref{Eq:FqF2scaing}. Then we can fit the values of $\beta_{q}$ and obtain the exponent $\nu$ by using Eq.~\eqref{Eq:betaqscaing}. The value of $\nu$ is found to be around $1.03\pm 0.01$, which is slightly larger than theoretical expectation, i.e. 1.0 in the Ising system~\cite{GLPRD,2DIsing}. It is caused by the finite event statistics and momentum resolution. It will give an upper limit to the number of maximum division cells and maximum order in real calculations.

  \begin{figure*}
     \centering
     \includegraphics[scale=0.91]{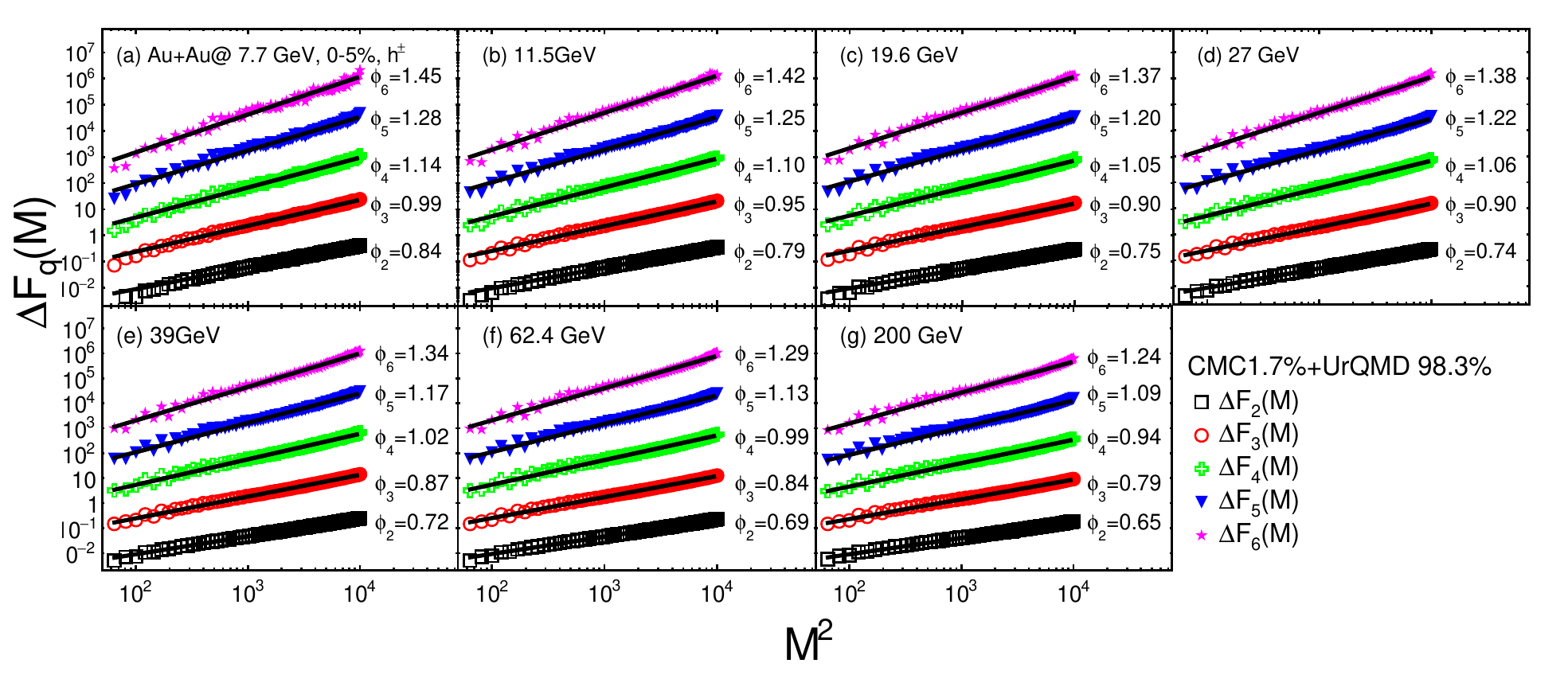}
      \caption{The correlator $\Delta F_{q}(M)$ ($q$=2-6) of charged particles as a function of $M^{2}$ in the most central (0-5\%) Au+Au collisions at $\sqrt{s_\mathrm{NN}}$ = 7.7-200 GeV from the UrQMD+CMC model with replacing fraction $\lambda$ = 1.7\%. The solid black lines represent the power-law fitting according to Eq.~\eqref{Eq:PowerLaw}.}
     \label{Fig:DeltaFqUrQMDCMC}
 \end{figure*}
 
\section{Apparent Intermittency in the UrQMD+CMC Model}
In the previous section, we have observed that the CMC model exhibits good intermittency behavior as expected. Nevertheless, it is a toy model which only produces momentum profiles of critically correlated particles and does not include the dynamical evolution as in heavy-ion collisions. One straightforward approach is to combine the CMC model with the UrQMD model, which aims to realize the presence of intermittency in heavy-ion collisions.  

To get the hybrid UrQMD+CMC model, part of the particles from the UrQMD model, which have already passed through the microscopic transport and final-state interactions, are substituted with those from the CMC simulation that have the same multiplicity and $p_{T}$ distributions. The replacing fraction is defined as~\cite{YigePLB}:
\begin{equation}
\lambda =\frac{N_{\rm CMC}}{N_{\rm UrQMD}},
 \label{Eq:fraction}
\end{equation}
\noindent where $N_{\rm CMC}$ is the number of CMC particles and $N_{\rm UrQMD}$ is the multiplicity in an original UrQMD event. To keep the $p_{T}$ distribution of the new UrQMD+CMC sample to be the same as that of the original UrQMD sample, we require the replacement to take place when $|p_{T}(\mathrm{CMC})-p_{T}(\mathrm{UrQMD})|<0.2$ (GeV/$c$) is satisfied. For a system with weak signal but strong background noises, as in the NA49 Si+Si collision, $\lambda$ is a small number. 

Figure~\ref{Fig:DeltaFqUrQMDCMC} depicts $\Delta F_{q}(M)$ as a function of $M^{2}$ in the most central (0-5\%) Au+Au collisions at $\sqrt{s_\mathrm{NN}}$ = 7.7-200 GeV from the UrQMD+CMC model with the replacing fraction $\lambda$ = 1.7\%. We observe that $\Delta F_{q}(M)$ ($q$=2-6) exhibit good power-law behaviors with increasing $M^{2}$ at various energies. It indicates that self-similar density fluctuations have been successfully incorporated into the UrQMD+CMC model. The solid black lines are the power-law fitting based on Eq.~\eqref{Eq:PowerLaw}. It is found that the intermittency indices of higher order $\Delta F_{q}(M)$ are larger than those of lower ones at various energies. 

 \begin{figure*}
     \centering
     \includegraphics[scale=0.91]{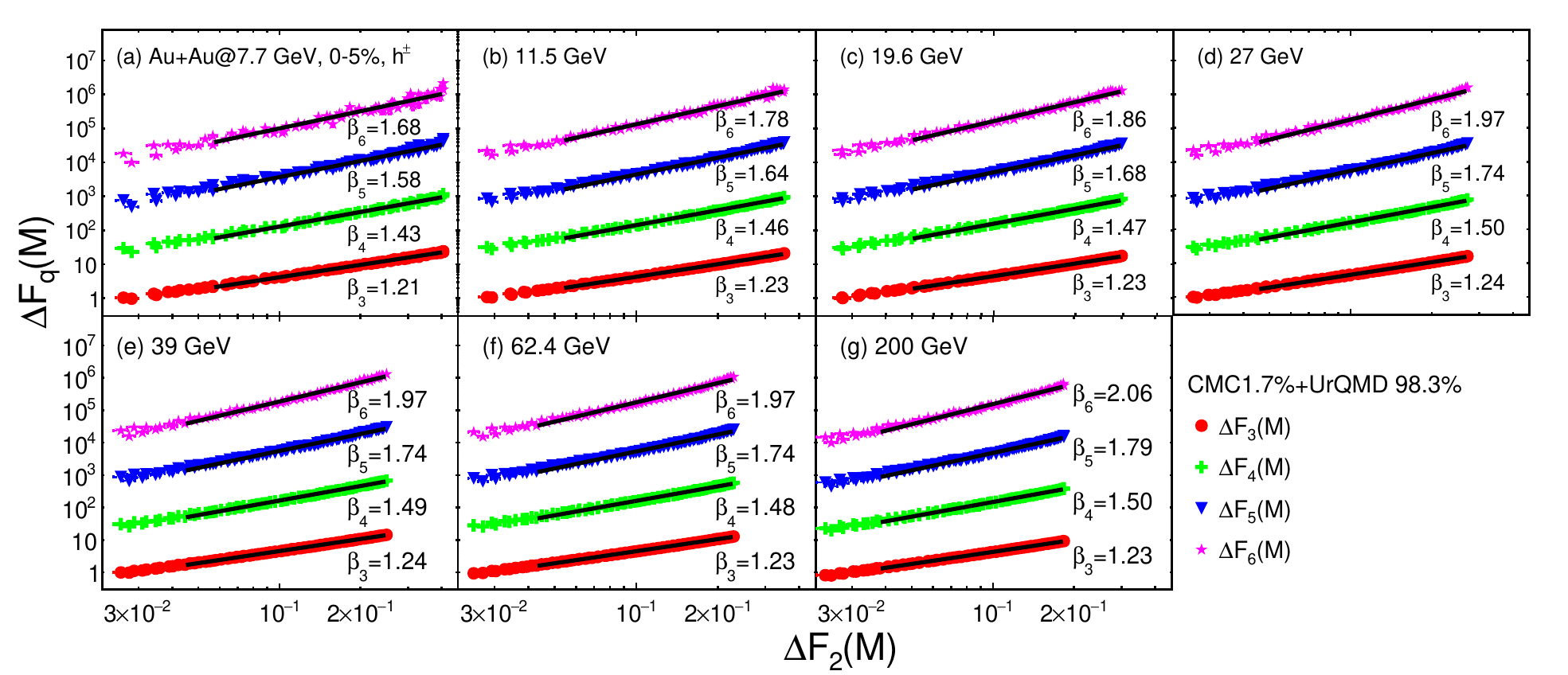}
      \caption{The higher-order $\Delta F_{q}(M)$ ($q$=3-6) as a function of $\Delta F_{2}(M)$ in the most central (0-5\%) Au+Au collisions at $\sqrt{s_\mathrm{NN}}$ = 7.7-200 GeV from the UrQMD+CMC model with $\lambda$ = 1.7\%. The solid black lines represent the power-law fitting according to Eq.~\eqref{Eq:FqF2scaing}}
     \label{Fig:DeltaFqF2UrQMDCMC}
 \end{figure*}
 
The STAR Collaboration at RHIC has recently measured intermittency of charged particles at BES-I energies. The preliminary result shows that the $\Delta F_{q}(M)/\Delta F_{2}(M)$ scaling is found in the most central Au+Au collisions~\cite{STARIntermittency,QMSTARIntermittency}. And this observation can not be described by the cascade UrQMD model. In the following, we will check whether the hybrid UrQMD+CMC model could reproduce the experimental measured scaling-law. 

In Fig.~\ref{Fig:DeltaFqF2UrQMDCMC}, we plot $\Delta F_{q}(M)$ ($q$=3-6) as a function of $\Delta F_{2}(M)$ calculated from UrQMD+CMC samples at seven RHIC BES-I energies with $\lambda$ = 1.7\%. In this case, $\Delta F_{q}(M)$ are found to obey good power-law scaling behaviors with increasing $\Delta F_{2}(M)$ at various energies, which agrees with what observed in the STAR experimental data~\cite{STARIntermittency,QMSTARIntermittency}. The solid black lines are the fitting according to the $F_{q}(M)/F_{2}(M)$ scaling in Eq.~\eqref{Eq:FqF2scaing}. The fitting range is chosen to be $M\in[30, 100]$, which is the same as that used in the STAR experimental analysis~\cite{STARIntermittency}. From these fitting, we can obtain $\beta_{q}$ and the scaling exponent $\nu$ by Eq.~\eqref{Eq:FqF2scaing} and \eqref{Eq:betaqscaing}, respectively.

 \begin{figure} 
     \includegraphics[scale=0.45]{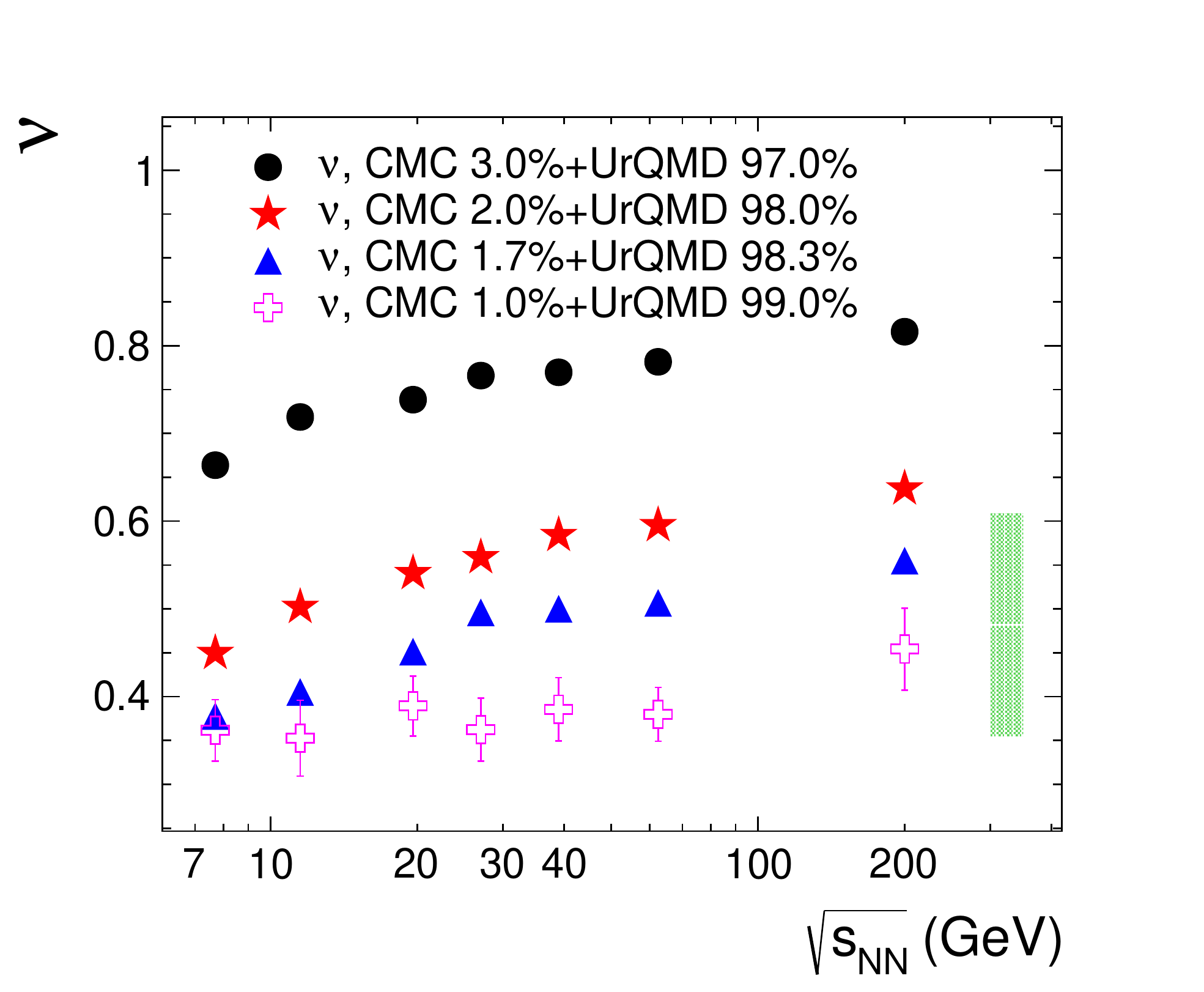}
     \caption{The energy dependence of the scaling exponent ($\nu$) in the most central (0-5\%) Au+Au collisions at $\sqrt{s_\mathrm{NN}}$ = 7.7-200 GeV from the UrQMD+CMC model with four selected replacing fractions. The green band illustrates the range of $\nu$ measured in the STAR experiment~\cite{STARIntermittency,QMSTARIntermittency}.}
     \label{Fig:nuenergy}
   \end{figure} 

In experiments, a possible intermittency signal may be shaded behind large background effects or other noises. First, finite size effects~\cite{PoberezhnyukPRC2020}, limited lifetime or critical slowing down of the system will restrict the growth of critical fluctuations in dynamic evolution of heavy-ion collision system~\cite{BerdnikovPRD2000}. Second, some trivial effects and experimental limitations, such as conservation law~\cite{BzdakPRC2013}, resonance decay and hadronic rescattering~\cite{ZhangPRC2020}, finite fluctuations inside the experimental acceptance~\cite{LingPRC2016,BzdakPRC2017,STARPRCMoment} as well as momentum resolution~\cite{SamantaJPG2021}, will weak or smear critical fluctuations. It is found that the observed power-law behavior in the NA49 experiment in Si+Si collisions can be reproduced by mixing 1\% of CMC particles with 99\% of random (uncorrelated) ones, indicating that the noise or background is indeed dominant in the experimental measurement~\cite{NA49EPJC}. It is meaningful to see how many percentages of intermittency signal could be related to the scaling behavior observed in the STAR experiment~\cite{STARIntermittency}. 

Fig.~\ref{Fig:nuenergy} illustrates the energy dependence of the scaling exponent $\nu$ in the most central (0-5\%) Au+Au collisions at $\sqrt{s_\mathrm{NN}}$ = 7.7-200 GeV from the UrQMD+CMC model with four different replacing fractions. We observe that all the $\nu$ calculated in the UrQMD+CMC model are smaller than 1.03, i.e. the value obtained in the pure CMC model. The reason is that large fraction of background particles from the UrQMD model fade the self-similar behavior. Furthermore, the values of $\nu$ get larger with higher replacing fractions. And they are found to monotonically increase with increasing $\sqrt{s_\mathrm{NN}}$ in all cases. This is due to more particles from the CMC model being included in the data samples with larger $\lambda$ or higher energies. The increase of the UrQMD particles in the mean time with energy has little effects on $\nu$ because the un-correlated background fluctuations have been subtracted by the mixed event method and the contribution to the value of $\nu$ is much smaller than that from the CMC particles. For comparison, the green band in the same figure denotes the range of $\nu$ (0.35-0.6) measured in the most central (0-5\%) Au+Au at $\sqrt{s_\mathrm{NN}}$ = 7.7-200 GeV from the preliminary results of the STAR experiment~\cite{STARIntermittency,QMSTARIntermittency}. We find that the calculated $\nu$ in the UrQMD+CMC model, with $\lambda$ chosen to be between 1\% and 2\%, fall in the experimentally measured range. Therefore, the UrQMD+CMC model can successfully reproduce the important scaling exponent measured by the STAR Collaboration. If infers that only 1-2 \% signal of intermittency could be related to the data sets from the STAR experiment, which is similar to value of $\lambda$ = 1\% in Si+Si collisions from the NA49 experiment~\cite{NA49EPJC}. 

The experimentally measured scaling exponent $\nu$ exhibits a non-monotonic behavior on beam energy and reaches a minimum around $\sqrt{s_\mathrm{NN}}$ = 20-30 GeV from the preliminary results of the STAR Collaboration~\cite{STARIntermittency,QMSTARIntermittency}. Our current hybrid UrQMD+CMC model cannot reproduce this non-monotonic energy dependence. It is due to a fixed replacing fraction $\lambda$ being used for various energies in this work. In a real experiment, the fraction of critical particles over background ones could depend on collision energy. This issue should be carefully taken into account in further study to investigate and understand the observed non-monotonic behavior at STAR.

\section{Summary and Outlook}
In this paper, we have investigated the intermittency of charged particles in Au+Au collisions at $\sqrt{s_\mathrm{NN}}$ = 7.7-200 GeV by using the cascade UrQMD, CMC, and hybrid UrQMD+CMC models, respectively. 

In the original UrQMD model, the values of SFMs are observed to overlap with those from the mixed events at various RHIC BES-I energies. Neither $\Delta F_{q}(M)/M$ nor $\Delta F_{q}(M)/\Delta F_{2}(M)$ scaling is valid when the background contributions from the mixed events are subtracted. It is consistent with the fact that there is no self-similar critical fluctuation in this model.

After including the same statistics, multiplicity, and transverse momentum distributions as those from the UrQMD samples, it is found that the calculated SFM from the CMC model is larger than that from the mixed events. Both the $\Delta F_{q}(M)/M$ and the $\Delta F_{q}(M)/\Delta F_{2}(M)$ scaling are satisfied in this case. It confirms that the scale-invariant intermittency behavior can be well simulated in the CMC model.

We incorporate self-similar density fluctuations generated from the CMC model into the event samples from the UrQMD to realize the presence of intermittency in the simulations. The hybrid UrQMD+CMC model exhibits good power-law dependence up to the sixth-order on the number of division cells in momentum space. The $\Delta F_{q}(M)/\Delta F_{2}(M)$ scaling is verified at various collision energies, which is consistent with the experimental results observed in STAR. As 1-2 \% signal of critical fluctuations from the CMC model is embedded, the energy dependence of the extracted scaling exponents show that the values are well within the experimentally measured range. It indicates that there only exists 1-2 \% of intermittency signal in the central Au+Au collisions from the STAR experiment. 

We would like to note that the intermittency signal may be easily ignored or missed in experimental measurements since both the NA49 and our work show that there only exist a few percentages of signal over background in current heavy-ion collisions. In order to understand these results, it is important to pick up the weak signal from background with strong noises. Recent studies show that a new computational technique~
\cite{Diakonos:2021epr} and machine learning method~\cite{YigePLB} may shed light on this direction. It is worthwhile to apply these new methods to experimental analysis to get a clean and reliable result and to learn the self-similar density fluctuation or intermittency behavior that could signal the second-order phase transition in the QCD phase diagram.

\section* {Acknowledgements}
This work is supported by the National Natural Science Foundation of China (Grants No. 12275102, No. 12122505, No. 11890711), National Key Research and Development Program of China (Grants No. 2020YFE0202002, No. 2022YFA1604900 and No. 2018YFE0205201) and the Fundamental Research Funds for the Central Universities (Grants No. CCNU220N003).

\bibliography{bib}		
\end{document}